\pdfoutput=1 

\documentclass{JINST}
\let\ifpdf\relax

\title{The GREAT3 Challenge}

\newcommand{\apjs}{ApJS}
\newcommand{\aj}{AJ}
\newcommand{\mnras}{MNRAS}
\newcommand{\aaps}{A\&AS}

\author{H.~Miyatake$^a$\thanks{Corresponding author.}~,
R.~Mandelbaum$^b$
and B.~Rowe$^c$
on behalf of the GREAT3 collaboration
\\
\llap{$^a$}Department of Astrophysical Sciences, Princeton University,\\
  4 Ivy Lane, Princeton, NJ 08544, USA\\
\llap{$^b$}McWilliams Center for Cosmology, Carnegie Mellon University,\\
  5000 Forbes Avenue, Pittsburgh, PA 15213, USA\\
\llap{$^c$}Department of Physics and Astronomy, University College London,\\
  Gower Street, London WC1E 6BT, UK\\
E-mail: \email{miyatake@astro.princeton.edu}}

\abstract{The GRavitational lEnsing Accuracy Testing 3 (GREAT3)
challenge is an image analysis competition that aims to test algorithms
to measure weak gravitational lensing from astronomical images. The
challenge started in October 2013 and ends 30 April 2014. The challenge focuses on testing the impact on weak lensing
measurements of realistically complex galaxy morphologies, realistic
point spread function, and combination of multiple different
exposures. It includes simulated ground- and space-based data. The
details of the challenge are described in [15], and
the challenge website and its leader board can be found at
http://great3challenge.info and
http://great3.projects.phys.ucl.ac.uk/leaderboard/, respectively.}


\begin{document}

\section{Introduction}\label{sec:introduction}
Precise cosmological measurements over the past decade enabled us to
establish the standard cosmological model, or $\Lambda$CDM model, which
revealed the fact that about 27\% of the energy of the Universe is
dark matter and about 68\% is dark
energy \cite{Planck:2013}. Although the existence of the dark components
is inferred by observations, we do not yet know what they actually
are. This is one of the most mysterious and profound questions in modern physics.

Weak gravitational lensing is the subtle deflection of light from
distant galaxies that is caused by massive structures in the Universe
between an observer and the galaxies. As a result, a galaxy in an
astronomical image is slightly distorted, or sheared, compared to its original
image. Weak lensing is a powerful tool to explore the dark components
since it enables us to trace the structure of the Universe including
dark matter. Weak lensing measurements also allow us to infer the
property of dark energy, since it affects the structure of
the Universe through its accelerated expansion and the light propagation
through its impact on the geometry of the Universe.

To place constraints on cosmological parameters, several weak lensing
surveys are being carried out or proposed, including ground-based
surveys such as the Kilo-Degree Survey\footnote{http://kids.strw.leidenuniv.nl} (KiDS; 1,500~deg$^2$, 25~mag,
ongoing), the Dark Energy Survey\footnote{http://www.darkenergysurvey.org} (DES; 5,000~deg$^2$, 25~mag,
ongoing), Hyper Suprime-Cam\footnote{http://www.naoj.org/Projects/HSC/}
(HSC; 1,400~deg$^2$, 26~mag, from March 2014), and the Large Synoptic Survey
Telescope (LSST; 20,000~deg$^2$, 27~mag, from 2022
\cite{LSST_Science_Collaboration:2009}), and space-based surveys such as
Euclid (15,000 deg$^2$, 23~mag, from 2020 \cite{Laureijs:2011}) and
Wide-Field Infrared Survey Telescope-Astrophysics Focused Telescope
Assets (WFIRST-AFTA; 2,000 deg$^2$, 27~mag, from 2023
\cite{Spergel:2013}).

Before forming astronomical images we actually analyze, photons from a
galaxy are scattered by atmosphere (for a ground-based telescope) and
telescope optics, and then pixelated on the detector. These effects
form the point-spread function (PSF). When measuring weak lensing, it is
important to correct for the PSF, which blurs and distorts galaxy images. PSFs
are usually estimated by using stars since they are the impulse response
of PSFs.

Currently cosmological weak lensing measurements are limited by
statistical uncertainty due to intrinsic galaxy shapes. However, as
survey volumes grow, controlling the systematic uncertainties stemming
from the PSF correction and other algorithmic difficulties
becomes more important. For example, ongoing surveys such as KiDS,
DES, and HSC require a systematic uncertainty of less than $1$\% of the
lensing signal, which should be reduced down to $\sim0.1$\% in future
surveys such as LSST, WFIRST-AFTA and Euclid \cite{Amara:2008}.

The third GRavitational lEnsing Accuracy Testing (GREAT3) challenge is a
public image analysis challenge to test and facilitate development of
weak lensing measurement algorithms. The GREAT3 challenge started in
October 2013 and will end on 30 April, 2014.

This paper is organized as follows. In section
\ref{sec:previous_challenges} we describe previous challenges and the
context for the GREAT3 challenge. We then describe a brief summary of
the GREAT3 challenge in section \ref{sec:GREAT3} and possible future
updates in section \ref{sec:future_updates}. For those who would like to
know more details of the challenge, please refer to
\cite{Mandelnaum:2013}.

\section{Previous Challenges and GREAT3 Goals}
\label{sec:previous_challenges} The history of lensing analysis
challenges via blind analysis of simulated data goes back to the Shear
Testing Programme 1 (STEP1) \cite{Heymans:2006} and STEP2
\cite{Massey:2007}. In STEP1, galaxies were modeled as a de Vaucouleurs
bulge and exponential disc profile, and PSFs were drawn from {\it
SkyMaker}\footnote{http://www.astromatic.net/software/skymaker}, which
takes into account realistic optical models and atmospheric turbulence.
In STEP2 they tried to make galaxies more realistic by representing
morphologies and substructures of galaxies using shapelets, i.e., linear
combination of 2-dimensional orthogonal basis functions
\cite{Refregier:2003, Bernstein:2002}. The PSFs and
weak lensing shears were not functions of position. They tried to create
realistic galaxy images by distributing objects in a field of view, so
that participants were required to detect objects by themselves and some
objects were actually blended.

The GREAT08 challenge \cite{Bridle:2009} simplified the challenge by
employing parametric galaxy/PSF models (bulge + disc profile for galaxies
and Moffat profile for PSFs) and gridded galaxy images (no blended
galaxies). These simplifications helped to identify problems of lensing analysis. The GREAT10 challenge
\cite{Kitching:2010} made the challenge more realistic by introducing a
shear field that varied across the field like a real, cosmological shear
field in the ``Galaxy Challenge''. They also introduced a ``Star Challenge''
in which the PSF has spatial variation. Thanks to these challenges, the
accuracy of weak lensing shape measurement was improved by factors of
2-3. The best methods submitted to GREAT10 have slightly less than 1\%
bias. They found a strong dependence on accuracy as a function of
signal-to-noise ratio,
and weak dependence on galaxy type and size.

The GREAT3 challenge focuses on testing the impact of the following
three effects on weak lensing measurement. First, GREAT3 addresses
how actual galaxy morphologies and substructures affect galaxy shape
estimates by using ``real'' galaxy images, i.e., without assuming
any specific model for galaxy light profiles. Second,
GREAT3 aims to test realistic PSF and its variation across the field of
view. GREAT3 simulates realistic PSFs based on the turbulent atmosphere
and an actual telescope design. Although the GREAT10 Star Challenge
judged participants on the accuracy of PSF reconstruction, the metric of
GREAT3 is the accuracy of shear field reconstruction to test the impact
on shear estimates directly. Third, GREAT3 tests the combination of multiple
different exposures. When measuring galaxy shapes, we often use multiple
different exposures with different dithers to obtain a higher
signal-to-noise ratio. There are two ways to combine these exposures:
adding them together to form a stacked image, and fitting the multiple
exposures simultaneously to treat each exposure separately. GREAT3 will
serve as a test of methods for combining multiple
exposures. Details of these aspects are described in the following
section.

\section{The GREAT3 Challenge}
\label{sec:GREAT3}
\begin{figure}[tbp] 
\centering
\includegraphics[width=0.8\textwidth]{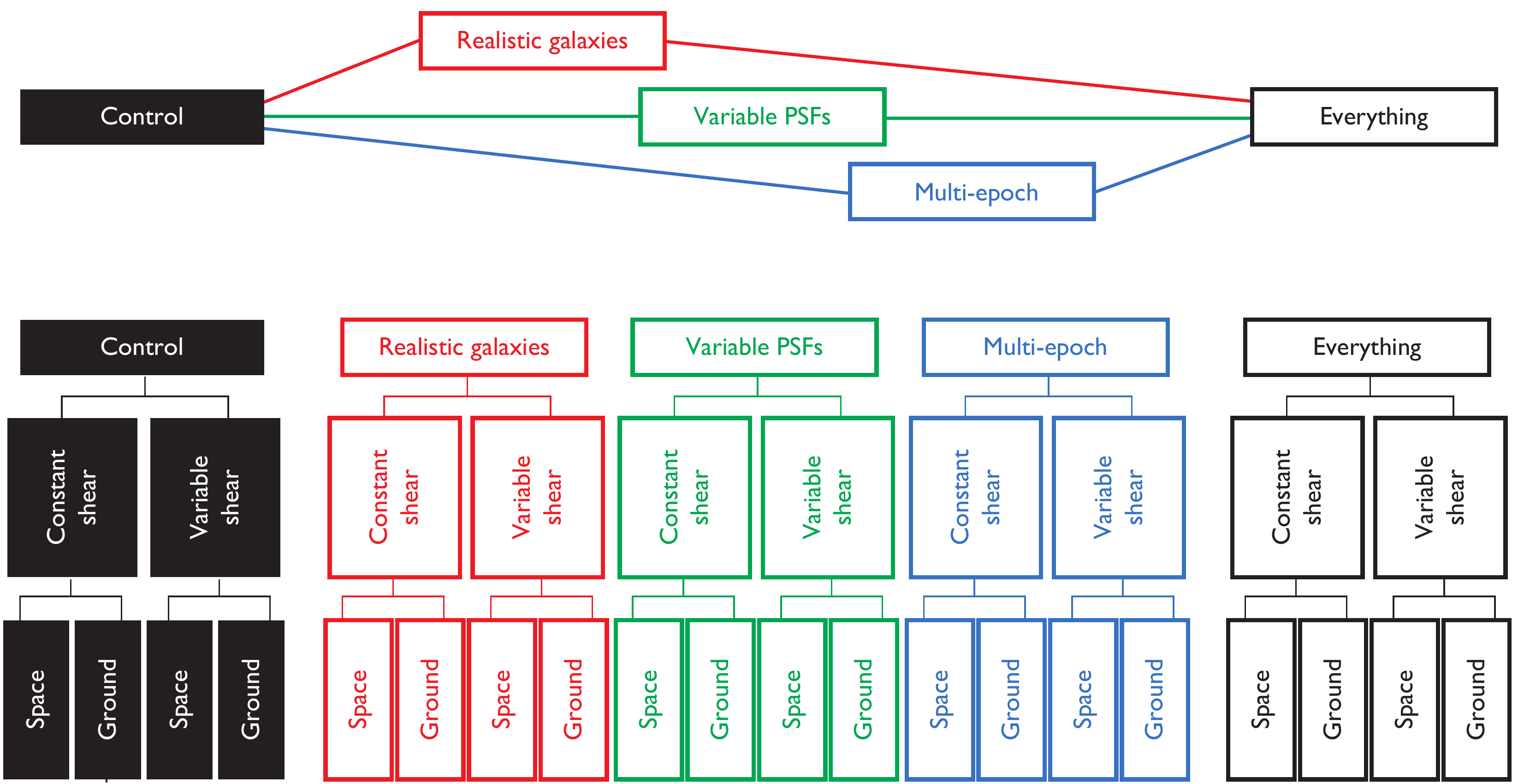}
\caption{The GREAT3 challenge structure. This figure is taken from \cite{Mandelnaum:2013}.}
\label{fig:challenge_structure}
\end{figure}

As described in section \ref{sec:previous_challenges}, GREAT3 focuses on
three aspects of image analysis, i.e., realistic galaxies, realistically
variable PSFs, and multiple different
exposures. Figure~\ref{fig:challenge_structure} shows the structure of
the GREAT3 challenge. The challenge consists of 5 experiments. The
control experiment does not have any of these three effects. The
realistic galaxy, variable PSF, and multi-epoch experiments are each
dedicated tests of a single effect. The `everything' experiment contains
all three effects in one experiment. Each experiment has branches with
constant shear and cosmologically-varying shear. For each shear
configuration, simulations for a space- and ground-based telescope are
prepared. Thus in total there are 20 branches.

The dominant statistical error for weak lensing shear measurement
comes from the dispersion in intrinsic galaxy shapes, or shape noise. We
design the challenge to cancel out the shape noise by introducing
90-degree rotated pairs in the constant shear branches and putting all
the shape noise into B-mode, as for GREAT10, in the variable shear
branches. This greatly increases the sensitivity of the tests for
systematic biases in shear measurements.

There are 200 images per branch. Each image has a grid of 100 $\times$
100 galaxies and represents 10 $\times$ 10 sq. degree field. Each
branch of the 14 branches with variable shear and/or variable PSF has 10
distinct fields that have different underlying shear correlation
functions and/or PSF variations. Each field has 20 subfields, and each
subfield corresponds to a single image. Participants should estimate the shear
correlation function by using the 20 subfields. On the other hand, for 6
branches with constant shear and constant PSF, shear values are
different image by image. In the following subsections, we describe the
details of these experiments except for the everything experiment. We
then explain the actual evaluations of submissions and ranking system.

\subsection{Control Experiment}
\label{sec:control_experiment} In the control experiment, galaxies are
generated based on a model that consists of a de Vaucouleurs bulge
profile and exponential disc profile or the model of a single component
S\'ersic profile. The details of the model fitting is described in
\cite{Lackner:2012}. The distributions of the model parameters, such as
size and ellipticity, are determined by the Hubble Space Telescope (HST)
COSMOS survey \cite{Koekemoer:2007, Scoville:2007a,
Scoville:2007b}. These distributions are not open to participants,
though the HST COSMOS data is publicly available.
The distribution of signal-to-noise ratio of galaxy images is also based
on the HST COSMOS survey. We set the lower limit of the signal-to-noise
ratio to $S/N=20$ using the optimal definition of
\cite{Bridle:2010}. There are several conventions for
signal-to-noise ratio. For
comparison with other conventions, we find that the one-sided 99\% lower
limits of our signal-to-noise ratio distribution actually corresponds to
$S/N$ values given by {\tt SExtractor} \cite{Bertin:1996} of $\simeq$10.0
(ground) and $\simeq$11.7 (space), which is comparable to the lower limit used
for real weak lensing analysis.

In the control experiment, PSFs are constant across an image. PSFs are
simulated in two parts: atmospheric PSF (only for ground-based
simulations) and optical PSF. For atmospheric PSFs, we use the
long-exposure limit atmospheric PSF model predicted by a Kolmogorov
model \cite{Fried:1965}. For a single exposure we draw the size and
ellipticity from their distributions estimated at the summit of Mauna
Kea by the LSST Image Simulator (PhoSim
\cite{Connolly:2010})\footnote{While the actual LSST exposure time
will be 15s, we choose random values of 60-180s since we try to simulate
``generic'' observations, for which this range is typical.}. For optical PSFs, we use a Zernike polynomial description of
wavefront errors up to order $j=11$ in the Noll ordering
\cite{Noll:1976}, including trefoil and third order spherical
aberration. The optical PSFs for space-based (ground-based) telescopes
are obtained by fitting the model to the WFIRST-AFTA prototype model
(the Dark Energy Camera early model). The PSF is given as a noiseless
image for each galaxy image\footnote{More exactly, 9 PSF images are
provided for a galaxy image. They have sub-pixel offsets except for the
lower left image, but all the images are based on the same underlying
PSF. However, they do not carry any additional information since images
are Nyquist sampled.}, so that participants should extract PSF
information from the image.
\subsection{Real Galaxy Experiment}
\begin{figure}[tbp] 
\centering
\includegraphics[width=0.4\textwidth]{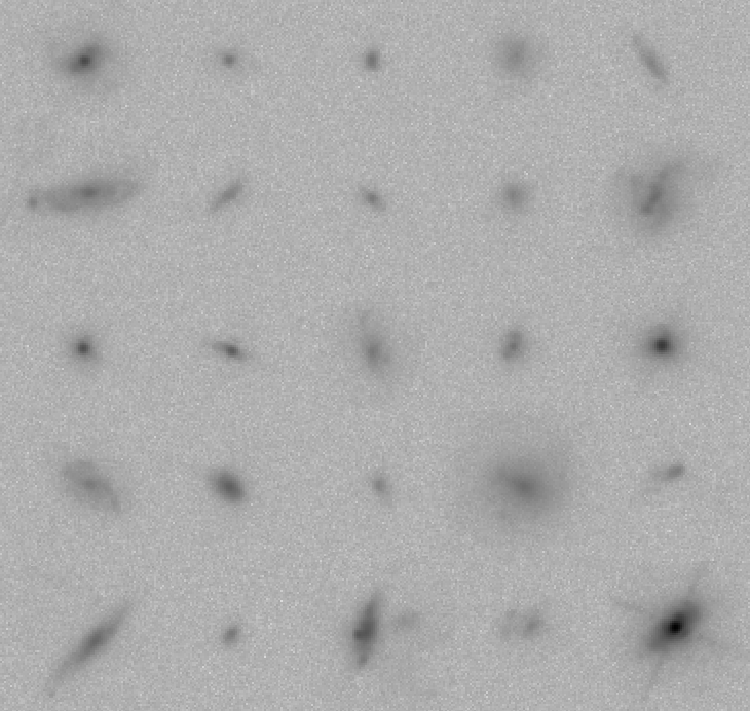}
\caption{An example of a realistic galaxy image simulation in
 GREAT3. Irregular galaxy morphology can be clearly seen.}
\label{fig:real_galaxy}
\end{figure}

For the real galaxy experiment, we use actual galaxy images
from the HST COSMOS survey. Following the procedure described in
\cite{Mandelbaum:2012}, the HST PSF is removed in Fourier space, the
weak lensing shear and magnification are applied, and then the target
PSF is applied. An example of realistic
galaxy image is shown in figure~\ref{fig:real_galaxy}. Note that
PSFs are given in the same way as described in section
\ref{sec:control_experiment}.

\subsection{Variable PSF Experiment}
\begin{figure}[tbp] 
\centering
\includegraphics[width=0.4\textwidth]{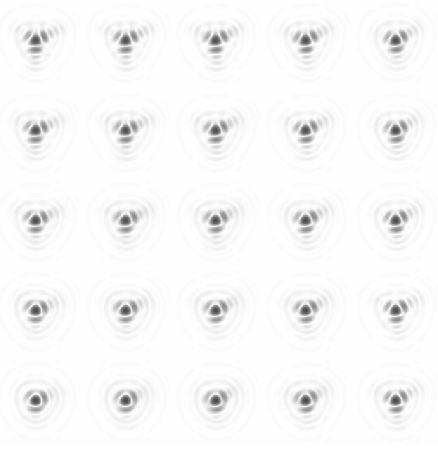}
\includegraphics[width=0.4\textwidth]{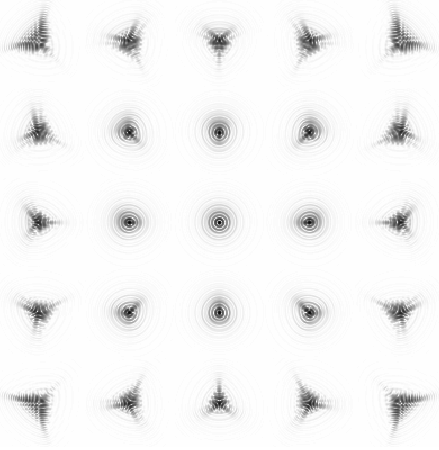}
\caption{An example of variable, optical  PSFs across the field of view. The left
 (right) panel shows simulations for a space-based telescope (a
 ground-based telescope). Both are shown on a logarithmic scale. This figure is taken from \cite{Mandelnaum:2013}.}
\label{fig:psf_variation}
\end{figure}

In the variable PSF experiment, PSFs vary across the field. In this
experiment we aim to test PSF interpolation across the field of view,
i.e., test if PSF information can be interpolated from star positions to
a galaxy position. The variation is drawn from PhoSim for atmospheric
PSFs and the actual optics model described above for optical PSFs. An
example of optical PSFs varying across the field is shown in
figure~\ref{fig:psf_variation}. Participants should extract the PSF
information from star images distributed across the field of view. Since
the simulated 10 $\times$ 10 sq. degree images are much larger than the
field-of-view of typical telescopes, images are divided into square
tiles, and a PSF model is simulated for each
tile. The star images have a signal-to-noise ratio distribution with
$S/N>50$ and a stellar density of about 2 arcmin$^{-2}$.  The galaxy
distribution is the same as for the control experiment.

\subsection{Multi-epoch Experiment}
In this experiment, by preparing 6 exposures with different PSFs and
dithering, we aim to find an optimal way to make use of the multiple
exposures, such as weighting exposures with good PSFs. The offsets
between exposures are provided, so that participants do not have to
infer them. The multi-epoch experiment employs the same galaxy
distribution as the control experiment. Note that the signal-to-noise
ratio distribution is recovered after all the exposures are
combined. The distribution is not known to participants, as in the
control experiment.

Data from space-based telescopes are often undersampled, i.e., not
Nyquist sampled, meaning that we need either to combine multiple
dithered exposures to obtain full information about PSFs, or
alternatively adopt some physically motivated model that can be fit to
stellar profiles to provide the lost (aliased) spatial information For
space branches in the multi-epoch experiment, we prepared non-Nyquist
sampled data to test this reconstruction. PSFs are given in the
same way as described in section \ref{sec:control_experiment}.

\subsection{Evaluations and Rankings}
Submissions and evaluations are done separately for each
branch. Participants compete within the leader board for each branch,
and are awarded points based on their best-ranked 5 branches. We award
1,000, 2,000, 4,000, 8,000, and 16,000
points for a fifth, fourth, third, second, and first-place finish, respectively. The
participant with the highest point total is the winner. The
first and second place winners will receive prizes.

The evaluations are done as follows. For constant shear branches,
participants submit mean shear value for each of the 200 fields (a
pre-submission code which calculates the mean shear based on each galaxy
shape and weight is available). A linear regression is performed to
provide estimates of $m_i$ and $c_i$ defined as $g^{\rm obs}_i -
g^{\rm true}_i = m_ig^{\rm true}_i+c_i$ where $i=1,2$ denotes shear
components , $g^{\rm obs}_i$ denotes the observed shear, and $g^{\rm
true}_i$ denotes the true input shear. The metric for constant shear
branches $Q_c$ is then defined by comparison of $m_i$ and $c_i$ to
target values of $m_{\rm target}=2\times10^{-3}$ and $c_{\rm
target}=2\times10^{-4}$, as required by the Euclid mission
\cite{Massey:2013}. We expect $Q_c\simeq1000$ ($700$) for space (ground)
when a method meets the target $m_i$ and $c_i$ and smaller $Q_c$ for
larger biases.

For variable shear branches, the evaluation involves shear correlation
functions, or more specifically, aperture mass dispersions (e.g.,
\cite{Schneider:1998}). A script for calculating aperture mass
dispersions is provided, though participants can also calculate the
statistics by themselves. The metric for variable shear $Q_v$ compares
the estimates and true values, which is designed to yield
$Q_v\simeq1000$ $(580)$ for space (ground) when a method meets the
target $m_i$ and $c_i$ and smaller $Q_v$ for larger biases. Higher scores for $Q_c$ and $Q_v$ are possible, due both to noise and the fact that methods may exceed target bias levels.

\section{Possible Future Updates}
\label{sec:future_updates}
Simulations in the GREAT3 challenge are generated by an open-source
image simulation software package,
GalSim\footnote{https://github.com/GalSim-developers/GalSim}, whose
details will be described in an upcoming paper \cite{Rowe:2014}. To test the three effects, we
simplify the GREAT3 simulation images. However, for example, using
GalSim, we are able to test star/galaxy separation by making images
where stars and galaxies are on the same image, blending issues (overlapping galaxy
images) by using non-gridded galaxies, and selection bias due to the
correlation between selection criteria and PSF/galaxy shapes.

Also there are possible extensions of GalSim to make simulations more
realistic. For instrument and detector specific effects, it might be
useful to include charge transfer inefficiency, non-linearity, and optics
and detector distortions. We will be able to test color gradients by
introducing wavelength-dependent effects on galaxies and PSFs. We can
test flexion measurement algorithms by adding higher-order distortions
across each galaxy.

\section{Conclusion}
The GREAT3 challenge is an image analysis competition to test and
facilitate weak lensing measurement algorithms. The challenge allows for
testing bias due to complex galaxy morphology by simulated images based
on real HST image, systematic uncertainties due to imperfect PSF
estimation of profiles and variations across the field by simulated PSFs
based on realistic atmospheric simulations and actual telescope optics
designs, and optimal analysis of multiple exposures by simulated
multi-epoch exposures. The challenge has started in October 2013 and
will close to participants on 30 April 2014. There are several
additional tests that can be done by using GalSim that is the core
simulation software for GREAT3 such as star/galaxy separation and galaxy
blending. There are several possible updates for GalSim such as
including detector effects, wavelength-dependent effects, and
flexion. The details of the challenge are described in
\cite{Mandelnaum:2013}.

\acknowledgments

This project was supported in part by NASA via the Strategic University
Research Partnership (SURP) Program of the Jet Propulsion Laboratory,
California Institute of Technology; and by the IST Programme of the
European Community, under the PASCAL2 Network of Excellence,
IST-2007-216886.  This article only reflects the authors' views.

HM acknowledges support from JSPS Postdoctoral Fellowships for Research
Abroad. RM was supported in part by program HST-AR-12857.01-A, provided
by NASA through a grant from the Space Telescope Science Institute,
which is operated by the Association of Universities for Research in
Astronomy, Incorporated, under NASA contract NAS5-26555. BR acknowledges
support from the European Research Council in the form of a Starting
Grant with number 240672. We also thank the Aspen Center for Physics and
the NSF Grant \#1066293 for their warm hospitality, where part of this
work was conducted.

\end{document}